\documentclass[conference]{IEEEtran}
\usepackage{verbatim}
\usepackage{array}
\usepackage{amsmath}
\usepackage{amsfonts}
\usepackage{amsthm}
\usepackage{amssymb}
\usepackage{multirow}

\newcommand{\Va}{\mathbf{a}}
\newcommand{\Vb}{\mathbf{b}}

\newcommand{\floor}[1]{\lfloor #1 \rfloor}

\newcommand{\N}{\mathcal{N}}

\newcommand{\R}{\mathcal{R}}
\newcommand{\Ra}{\mathcal{R}(\eta)}
\newcommand{\ket}[1]{|#1\rangle}

\newcommand{\weight}[1]{\ensuremath{\mathrm{w}\left(#1\right)}}
\newcommand{\symp}[2]{\ensuremath{\left\langle #1,#2 \right\rangle}}
\newcommand{\transpose}[1]{\ensuremath{#1^{\mathrm{T}}}}
\newcommand{\high}{\rule{0pt}{1.1em}}

\newcommand{\centraliser}[1]{\ensuremath{\overline{#1}}}

\newtheorem{theorem}{Theorem}[section]
\newtheorem{lemma}[theorem]{Lemma} \newtheorem{claim}[theorem]{Claim}
\newtheorem{definition}[theorem]{Definition}
\newtheorem{corollary}[theorem]{Corollary}
\newtheorem{proposition}[theorem]{Proposition}
\usepackage{pgf}
\usepackage{tikz}

\begin{document}

\title{Quantum Cyclic Code of length dividing $p^{t}+1$}
\author{\IEEEauthorblockN{Sagarmoy Dutta}
  \IEEEauthorblockA{Dept of Computer Science and Engineering\\
  Indian Institute of Technology Kanpur\\
  Kanpur, UP, India, 208016\\
  {\tt sagarmoy@cse.iitk.ac.in}}
  \and
  \IEEEauthorblockN{Piyush P Kurur}
  \IEEEauthorblockA{Dept of Computer Science and Engineering\\
  Indian Institute of Technology Kanpur\\
  Kanpur, UP, India, 208016\\
  {\tt ppk@cse.iitk.ac.in}}
  and\\
  Max-Planck Institut f\"ur Informatik\\
  Campus E1 4, 66123, Saarbr\"ucken, Germany \\
  }
\date{} \maketitle

\newcommand{\etal}[0]{\emph{et al}}

\begin{abstract}

  In this paper, we study cyclic stabiliser codes over $\mathbb{F}_p$
  of length dividing $p^t+1$ for some positive integer $t$. We call
  these $t$-Frobenius codes or just Frobenius codes for short. We give
  methods to construct them and show that they have efficient decoding
  algorithms.

  An important subclass of stabiliser codes are the linear stabiliser
  codes. For linear Frobenius codes we have stronger results: We
  completely characterise all linear Frobenius codes. As a
  consequence, we show that for every integer $n$ that divides $p^t+1$
  for an odd $t$, there are no linear cyclic codes of length $n$. On
  the other hand for even $t$, we give an explicit method to construct
  all of them. This gives us a many explicit example of Frobenius
  codes which include the well studied Laflamme code.

  We show that the classical notion of BCH distance can be generalised
  to all the Frobenius codes that we construct, including the
  non-linear ones, and show that the algorithm of Berlekamp can be
  generalised to correct quantum errors within the BCH limit. This
  gives, for the first time, a family of codes that are neither CSS
  nor linear for which efficient decoding algorithm exits.

  The explicit examples that we construct are summarised in
  Table~\ref{tab:explicit-examples-short} and explained in detail in
  Tables~\ref{tab:explicit-examples-2} (linear case) and
  \ref{tab:explicit-examples-3} (non-linear case).

\end{abstract}

\section{Introduction}

Successful implementation of quantum computing requires handling
errors that occur while processing, storing and communicating quantum
information. Good quantum error correcting codes are therefore a key
technology in the eventual building of quantum computing devices,
besides, perhaps more importantly, their theory provide some elegant
mathematics. An important class of codes are the stabiliser
codes~\cite{gottesman:1996:error}, which not only captured the
isolated examples constructed
earlier~\cite{Sho95,Ste96a,BDSW96,LMPZ96}, but built a solid
foundation for subsequent
works~\cite{calderbank98quantum,Ashikhmin2000nonbinary,arvind2003family}.

Constructing stabiliser codes require handling the slightly
non-standard symplectic inner product. The CSS
construction~\cite{calderbank96css,steane96css} gives one elegant and
natural way, albeit with some loss of generality, to handle this
difficulty. For this one needs a self-dual code classical code, or
more generally two classical codes one contained in the dual of the
other, thereby reusing the intuition built for classical codes.
Another approach to the problem, again with some loss of generality,
is to look at linear stabiliser
codes~\cite{calderbank98quantum}. Linear stabiliser codes can also be
characterised as linear classical codes over a quadratic extension of
the base field~\cite[Theorem 3]{calderbank98quantum}~\cite[Lemma
18]{ketkar2005nonbinary} which are Hermitian self-dual.

In this article, we study mainly cyclic stabiliser codes. Cyclic
codes, being well studied classically, have recently been studied in
detail~\cite{calderbank98quantum,thangaraj2001quantumcyclic,salah2006quantumBCH,ketkar2005nonbinary},
mostly from the perspective of either self dual codes or Hermitian
self dual codes. We explore another approach to simplify the
symplectic condition, namely, we restrict our attention of cyclic
codes of length dividing $p^t +1$ over $\mathbb{F}_p$.

\subsubsection*{Our contribution}

In this article, we focus on cyclic stabiliser codes over the field
$\mathbb{F}_p$ whose lengths divide $p^t+1$, for some positive integer
$t$. We call such codes $t$-\emph{Frobenius codes}, or just
\emph{Frobenius codes}, because of the key role played by the
Frobenius automorphism. Restricting to such lengths, while
constraining, is not that bad, as there is a healthy, i.e. almost
linear, density of such lengths. In bargain, we get a
simpler formulation of the isotropy condition, which helps in the
analysis of these codes considerably. Furthermore, this simplicity of
the isotropic condition allows us to extend the notion of BCH distance
for these codes and give efficient decoding algorithms. Since none of
the codes that we construct are CSS --- all our codes are uniquely
cyclic (See Section~\ref{sec:quantum-cyclic-codes} for a definition)
and by Proposition~\ref{prop:uniquely-cyclic-is-non-css} are not CSS
--- and some of them are non-linear, this gives a family of codes for
which efficient decoding algorithms were not known before.

We study the subfamily of \emph{linear} Frobenius codes in detail and
completely characterise them
(Theorems~\ref{thm:linear-necessary-condition}
and~\ref{thm:main-construction}). This has two consequence, one
negative and another positive. Firstly, over $\mathbb{F}_p$, we show
that there are no $t$-Frobenius linear codes when $t$ is odd
(Corollary~\ref{cor:nonexistance-power-odd}). This is a somewhat
serious limitation of linear cyclic codes as the density of such
lengths $n$ seems to be almost linear. Moreover, this impossibility is purely Galois theoretic
unlike other known restriction that arise from sphere packing bounds
or linear programming bounds.

On the positive side, the characterisation of linear Frobenius codes
gives us ways to explicitly construct examples of linear Frobenius
codes of lengths $p^{2t}+1$. Again, since the density of such lengths
are also healthy, this technique give sizable number of explicit
examples including the well studied Laflamme
code. Table~\ref{tab:explicit-examples-short} give such examples for
$p=2$ and lengths less than 100.





\section{Preliminaries}\label{sec:prelims}

We give a brief overview of the notation used in this paper. For a
prime power $q = p^k$, $\mathbb{F}_q$ denotes the unique finite field
of cardinality $q$. The product $\mathbb{F}_p^n$ is a vector space
over the finite field $\mathbb{F}_p$ and an element
$\mathbf{a}=(a_1,\ldots,a_n)^\mathrm{T}$ in it is thought of as a column
vectors.  Fix a $p$-dimensional Hilbert space $\mathcal{H}$. An
orthonormal basis for $\mathcal{H}$ is of cardinality $p$. Fix one
such basis and denote it by $\{ \ket{a} | a \in \mathbb{F}_p \}$.  As
is standard in quantum computing, for an element
$\mathbf{a}=(a_1,\ldots,a_n)^\mathrm{T}$ in $\mathbb{F}_p^n$,
$\ket{\mathbf{a}}$ denotes the tensor product
$\ket{a_1}\otimes\ldots\otimes\ket{a_n}$.  The set $\{
\ket{\mathbf{a}} | \mathbf{a} \in \mathbb{F}_p^n \}$ forms a basis for
the $n$-fold tensor product $\mathcal{H}^{\otimes^n}$. A
\emph{quantum code} over $\mathbb{F}_p$ of \emph{length} $n$ is a
subspace of the tensor product $\mathcal{H}^{\otimes^n}$. There is by
now a significant literature on quantum codes
\cite{knill2000theory,gottesman:1996:error,calderbank98quantum}.

Let $\zeta$ denote the primitive $p$-th root of unity
$\exp{\frac{2\pi\iota}{p}}$.  For $\Va$ and $\Vb$ in $\mathbb{F}_p^n$,
define the operators $U_{\mathbf{a}}$ and $V_{\mathbf{b}}$ on
$\mathcal{H}^{\otimes^n}$ as $U_\mathbf{a} \ket{\mathbf{x}} =
\ket{\mathbf{x} + \mathbf{a}}$ and $V_\mathbf{b} \ket{\mathbf{x}} =
\zeta^{\mathbf{b}^\mathrm{T} \mathbf{x}} \ket{\mathbf{x}}$
respectively. The operator $U_\mathbf{a}$ can be thought of as a
\emph{position error} and $V_\mathbf{b}$ as a \emph{phase error}. In a
quantum channel, both position errors and phase errors can occur
simultaneously. These are captured by the \emph{Weyl} operators
$U_\mathbf{a} V_\mathbf{b}$.

For elements $\mathbf{a}$ and $\mathbf{b}$ of the vector space
$\mathbb{F}_p^n$ the \emph{joint weight}
$\weight{\mathbf{a},\mathbf{b}}$ is the number of positions $i$ such
that either $a_i$ or $b_i$ is not zero. The \emph{weight} of the Weyl
operator $U_\mathbf{a}V_\mathbf{b}$ is the joint weight
$\weight{\mathbf{a},\mathbf{b}}$. Occurrence of a quantum error at
some $t$ positions is modelled as the channel applying an unknown Weyl
operator $U_\mathbf{a}V_\mathbf{b}$ of weight $t$ on the transmitted
message.

An important class of quantum codes are the class of stabiliser
codes~\cite{gottesman:1996:error}. One can study stabiliser codes by
studying the isotropic sets under the \emph{symplectic inner product}.
For any two vectors $\mathbf{u}=(\mathbf{a},\mathbf{b})$ and
$\mathbf{v}=(\mathbf{c},\mathbf{d})$ of
$\mathbb{F}_p^{n}\times\mathbb{F}_p^n$, define the \emph{symplectic
  inner product} $\symp{\mathbf{u}}{\mathbf{v}}$ as the scalar
$\transpose{\mathbf{a}}\mathbf{d} - \transpose{\mathbf{b}}\mathbf{c}$
of $\mathbb{F}_p$.  A subset $S$ of $\mathbb{F}_p^{2n}$ is called
\emph{totally isotropic}~\cite{calderbank98quantum}, or just
\emph{isotropic}, if for any two elements $\mathbf{u}$ and
$\mathbf{v}$ of $S$, $\symp{\mathbf{u}}{\mathbf{v}} = 0$. 

Isotropic subspaces of $\mathbb{F}_p^{2n}$ are closely related to
stabiliser codes. Calderbank
\etal~\cite{calderbank97orthogonal,calderbank98quantum} were the first
to study this relation when the underlying field is
$\mathbb{F}_2$. Later, this was generalised to arbitrary
fields~\cite{Ashikhmin2000nonbinary,arvind2003family}. We summaries
these results in a form convenient for our purposes.
\begin{theorem}[\cite{calderbank97orthogonal,Ashikhmin2000nonbinary,
    arvind2003family}]%
  \label{thm:iso-stab-connection}
  Let $S$ be a isotropic subspace of $\mathbb{F}_p^{2n}$ for some
  positive integer $n$. Let $\omega$ be either the primitive $p$-th
  root of unity $\exp{\frac{2\pi \iota}{p}}$ or $\sqrt{-1}$, depending
  on whether $p$ is odd or even respectively.  Then, the subset
  $\mathcal{S} = \{ \omega^{\Va^T \Vb} U_\Va V_\Vb | (\Va,\Vb) \in S
  \}$ of unitary operators forms an Abelian group. Furthermore, the
  set of vectors invariant under the operators in $\mathcal{S}$ forms
  a quantum stabiliser code and the operator $P = \sum_{U \in
    \mathcal{S}} U$ is the projection to it.
\end{theorem}

Let $S$ be a subspace of $\mathbb{F}_p^{2n}$. By the
\emph{centraliser} of $S$, denoted by $\centraliser{S}$, we mean the
subspace of all $\mathbf{u}$ in $\mathbb{F}_p^{2n}$, such that
$\symp{\mathbf{u}}{\mathbf{v}} = 0$, for all $\mathbf{v}$ in $S$. We
have the following theorem on the error correcting properties of the
stabiliser codes.

\begin{theorem}[\cite{calderbank97orthogonal,Ashikhmin2000nonbinary, arvind2003family}]
  \label{thm:isotropic-main-properties}
  Let $S$ be a isotropic subspace of $\mathbb{F}_p^{2n}$ and let
  $\mathcal{C}$ be the associated stabiliser code. Then the dimension
  of the subspace $S$ is at most $n$. If $S$ has dimension $n - k$ for
  some $k> 0$ then the centraliser $\centraliser{S}$, as a vector
  space over $\mathbb{F}_p$, is of dimension $n+k$ and the code
  $\mathcal{C}$, as a Hilbert space, is of dimension $p^{k}$.
  Furthermore, if the minimum weight
  $\textrm{min}\{\weight{\mathbf{u}} | \mathbf{u} \in \centraliser{S}
  \setminus S \}$ is $d$ then $\mathcal{C}$ can detect up to $d-1$
  errors and correct up to $\floor{\frac{d - 1}{2}}$ errors.
\end{theorem}

Let $\mathcal{C}$ be a stabiliser code associated with an $n-k$
dimensional totally isotropic subspace $S$ of $\mathbb{F}_p^{2n}$. By
the \emph{stabiliser dimension} of $\mathcal{C}$ we mean the integer
$k$. Similarly, we call the weight $\textrm{min}\{ \weight{\mathbf{u}}
| \mathbf{u} \in \centraliser{S} \setminus S \}$ the \emph{distance}
of $\mathcal{C}$.  In this context, recall that the stabiliser code
associated to the isotropic set $S$ is called \emph{$\delta$-pure}, if the
minimum of the joint weights of non-zero elements of the centraliser
$\centraliser{S}$ is $\delta$. It follows from
Theorem~\ref{thm:isotropic-main-properties} that a $\delta$-pure code is of
distance at least $\delta$. A stabiliser code over $\mathbb{F}_p$ of length
$n$, stabiliser dimension $k$ and distance $\delta$ is called an
$[[n,k,\delta]]_p$ code.

\section{Quantum Cyclic codes}\label{sec:quantum-cyclic-codes}

In this section we define quantum cyclic codes and study some of its
properties. Fix a prime $p$ and a positive integer $n$ coprime to $p$
for the rest of the section. Let $N$ denote the right shift operator
over $\mathbb{F}_p^n$, i.e.  the operator that maps $\mathbf{u} =
(u_1,\ldots,u_n)$ to $(u_n,u_1,\ldots,u_{n-1})$.  Consider the unitary
operator $\N$ defined as $\N \ket{\mathbf{u}} = \ket{N \mathbf{u}}$.
Recall that a classical code over $\mathbb{F}_p$ is cyclic if for all
code words $\mathbf{u}$, its right shift $N\mathbf{u}$ is also a code
word. Motivated by this definition, we have the following definition
for quantum cyclic codes.

\begin{definition}\label{def_qcc} 
  A quantum code $\mathcal{C}$ is \emph{cyclic} if for any vector
  $\ket{\psi}$ in $\mathcal{C}$, the vector $\N \ket{\psi}$ is in
  $\mathcal{C}$.
\end{definition}

Let $S$ be a subspace of $\mathbb{F}_p^n \times \mathbb{F}_p^n$. We
say that $S$ is \emph{simultaneously cyclic} if for all $(\Va,\Vb)$ in
$S$, $(N\Va,N\Vb)$ is also in $S$. Stabiliser codes with simultaneously
cyclic isotropic sets were first studied by Calderbank
\etal~\cite[Section 5]{calderbank98quantum} and was taken as the
definition of cyclic codes in subsequent
works~\cite{thangaraj2001quantumcyclic,salah2006quantumBCH,ketkar2005nonbinary}. In
this context, we show that for stabiliser codes, simultaneous
cyclicity and our definition of cyclicity coincide.

\begin{proposition}\label{prop:stab-cyclic} 
  An isotropic subset of $\mathbb{F}_p^{n}\times\mathbb{F}_p^n$ is
  simultaneously cyclic if and only if the associated stabiliser code
  is cyclic.
\end{proposition} 
\begin{IEEEproof} 
  For a code $\mathcal{C}$ with projection operator $P$, it is easy to
  verify that $\mathcal{C}$ is cyclic if and only if $\N^\dag P \N =
  P$.  Let $S$ be an isotropic subset of
  $\mathbb{F}_p^{n}\times\mathbb{F}_p^n$ and let $\mathcal{C}$ be the
  associated stabiliser code.

  Assume that $\mathcal{C}$ is cyclic.  From
  Theorem~\ref{thm:iso-stab-connection}, the projection operator to
  $\mathcal{C}$ is given by $P = \sum_{(\Va,\Vb) \in S}
  \alpha_{\Va,\Vb} U_\Va V_\Vb$, where $\alpha_{\Va,\Vb} =
  \omega^{\Va^T \Vb}$. Notice that $\N^\dag U_\Va V_\Vb \N = U_{N \Va}
  V_{N \Vb}$. Therefore, the set $S$ should be simultaneously cyclic,
  otherwise the support of $\N^\dag P \N$ will not match with that of
  $P$.
  
  Conversely, if $S$ is simultaneously cyclic, then we have $(N \Va, N
  \Vb)$ is in $S$ for all $(\Va,\Vb)$ in $S$. The inverse of the shift
  operation $N$ is just $\transpose{N}$. Therefore, $\Va^T N^T N \Vb =
  \Va^T \Vb$ and hence the scalars $\alpha_{\Va,\Vb}$ are also
  preserved. Thus, $\N^\dag P \N = P$ and as a result, $\mathcal{C}$
  is cyclic.
\end{IEEEproof}

%
%

Let $\R$ denote the cyclotomic ring $\mathbb{F}_p[X]/X^n -1$ of
polynomials modulo $X^n -1$. When dealing with cyclic codes, it is
often convenient to think of vectors of $\mathbb{F}_p^n$ as
polynomials in $\R$ by identifying the vector
$\mathbf{a}=(a_0,\ldots,a_{n-1})$ with the polynomials $a(X) = a_0 +
\ldots + a_{n-1} X^{n-1}$. We use the bold face Latin letter, for
example $\mathbf{a}$, $\mathbf{b}$ etc, to denote vectors and the
corresponding plain face letter, $a(X)$, $b(X)$ respectively, for the
associated polynomial.  Recall that, classical cyclic codes are ideals
of this ring $\R$. In the ring $\R$, the polynomial $X$ has a
multiplicative inverse namely $X^{n-1}$. Often, we write $X^{-1}$ to
denote this inverse.  Notice that for any two vectors $\mathbf{a}$ and
$\mathbf{b}$ in $\mathbb{F}_p^n$, if $a(X)$ and $b(X)$ denote the
corresponding polynomials in $\R$, then the coefficient of $X^k$ in
the product $a(X)b(X^{-1})\mod X^n -1$ is the inner product
$\transpose{\mathbf{a}}N^k\mathbf{b}$, where $N$ is the right shift
operator. An immediate consequence is the following.

\begin{proposition}\label{prop:isotropy-polynomials}
  Let $S$ be a simultaneously cyclic subset of $\mathbb{F}_p^n \times
  \mathbb{F}_p^n$. Then $S$ is isotropic if and only if for any two
  elements $\mathbf{u} = (\mathbf{a},\mathbf{b})$ and $\mathbf{v} =
  (\mathbf{c},\mathbf{d})$, the corresponding polynomials satisfy the
  condition
  \[
  b(X)c(X^{-1}) - a(X)d(X^{-1}) = 0 \mod X^n -1.
  \]
\end{proposition}

Let $S$ be a simultaneously cyclic subspace of $\mathbb{F}_p^{n}\times
\mathbb{F}_p^n$.  Define $A$ and $B$ to be the projections of $S$ onto
the first and last $n$ coordinates respectively, i.e. $A=\{ \mathbf{a}
| (\mathbf{a},\mathbf{b}) \in S \}$ and $B = \{ \mathbf{b} |
(\mathbf{a},\mathbf{b}) \in S \}$. Since $S$ is simultaneously cyclic,
$A$ and $B$ are cyclic subspaces of $\mathbb{F}_p^n$ and hence are
ideals of the ring $\R$. Let $g(X)$ be the factor of $X^n - 1$ that
generates $A$. Since $g(X)$ is an element of $A$, there exists a
polynomial $f(X)$ in $\R$ such that $(\mathbf{g},\mathbf{f}) \in
S$. If this $\mathbf{f}$ is unique then we say that $S$ is
\emph{uniquely cyclic} and call the pair $(g(X),f(X))$ of polynomials,
a \emph{generating pair} for $S$. We have the following proposition.

\begin{proposition}\label{prop:uniquely-cyclic}
  A simultaneously cyclic subspace $S$ of $\mathbb{F}_p^n \times
  \mathbb{F}_p^n$ is uniquely cyclic if and only if for every element
  $(0,\mathbf{a})$ in $S$, $\mathbf{a} = 0$. If $S$ is uniquely cyclic
  generated by the pair $(g,f)$, then every element of $S$ is of the
  form $(a g, a f)$ for some $a(X)$ in $\mathbb{F}_p[X]/X^n -1$.
\end{proposition}

For a CSS code, the underlying isotropic set $S$ is a product
$C_1\times C_2$ of two $n$-length classical codes over $\mathbb{F}_p$
In particular, elements $(\mathbf{a},0)$ and $(0,\mathbf{b})$ for
$\mathbf{a}$ and $\mathbf{b}$ in $C_1$ and $C_2$ respectively belong
to $S$.  Therefore, we have the following proposition as a
consequences of Proposition~\ref{prop:uniquely-cyclic}.

\begin{proposition}\label{prop:uniquely-cyclic-is-non-css}
  Any uniquely cyclic stabiliser code is not CSS unless it is of
  distance $1$.
\end{proposition}

For uniquely cyclic codes the isotropy condition in
Proposition~\ref{prop:isotropy-polynomials} can be simplified as
follows.

\begin{proposition}\label{prop:isotropy-uniquely-cyclic}
  Let $S$ be a simultaneously cyclic subspace of $\mathbb{F}_p^n
  \times \mathbb{F}_p^n$ with generating pair $(g,f)$. Then $S$ is
  isotropic if and only if $g(X)f(X^{-1})=g(X^{-1})f(X)$ modulo $X^n
  -1$. Moreover, any pair $(a,b)$ belongs to $\centraliser{S}$ if and
  only if $g(X)b(X^{-1})=a(X^{-1})f(X)$ modulo $X^n-1$.
\end{proposition}

Consider a quadratic extension $\mathbb{F}_{p^2}=\mathbb{F}_p(\eta)$
of $\mathbb{F}_p$ obtained by adjoining a root $\eta$ of some quadratic
irreducible polynomial over $\mathbb{F}_p$. Identify the product
$\mathbb{F}_p^n \times \mathbb{F}_p^n$ with the the vector space
$\mathbb{F}_{p^2}^n$ by mapping a pair of vectors $(\mathbf{a},
\mathbf{b})$ to the vector $\mathbf{a} + \eta \mathbf{b}$. Similarly
for the cyclotomic ring $\R$, identify the product ring $\R \times \R$
with the cyclotomic ring $\R(\eta) = \mathbb{F}_{p^2}[X]/X^n -1$.  Let
$S$ be any isotropic subspace of $\mathbb{F}_p^n \times
\mathbb{F}_p^n$. The associated stabiliser code $\mathcal{C}_S$ is
said to be \emph{linear}~\cite{calderbank98quantum} if $S$ under the
above identification is a subspace of $\mathbb{F}_{p^2}^n$.  Isotropic
subspaces of $\mathbb{F}_p^n \times \mathbb{F}_p^n$ associated to
linear stabiliser codes are classical cyclic codes of length $n$ over
$\mathbb{F}_p(\eta)$. Thus the following proposition follows.

\begin{proposition}\label{prop:centraliser-ideal-connection}
  Let $S$ be an isotropic simultaneously cyclic subspace of the product
  $\mathbb{F}_p^n \times \mathbb{F}_p^n$. The associated stabiliser
  code $\mathcal{C}_S$ is linear if and only if $S$ is an ideal of the
  cyclotomic ring $\mathbb{F}_{p^2}[X]/X^n -1$. Furthermore, if
  $\mathcal{C}_S$ is linear then the centraliser $\centraliser{S}$ is
  also an ideal of $\mathbb{F}_{p^2}[X]/X^n -1$.
\end{proposition}

It follows from the theory of classical codes that both $S$ and
$\centraliser{S}$ are ideals generated by factors of $X^n -1$ over
$\mathbb{F}_{p^2}$. In this context, we make the following definition.

\begin{definition}[BCH distance]
  Let $g(X)$ be a factor of the polynomial $X^n - 1$ over the field
  $\mathbb{F}_q$, $n$ coprime to $q$. The \emph{BCH distance} of the
  polynomial $g(X)$ is the largest integer $d$ such that the
  consecutive distinct powers $\beta^{\ell}$,$\beta^{\ell+1},\ldots,
  \beta^{\ell + d -2}$ are roots of $g$, for some primitive $n$-th
  root $\beta$.
\end{definition}

Recall that, the distance of a classical cyclic code is at least the
BCH distance of its generating polynomial. In the setting of
stabiliser codes, the distance is related to the minimum joint weight
of elements of $\centraliser{S}$
(Theorem~\ref{thm:isotropic-main-properties}). Motivated by this
analogy, we define the BCH distance of linear stabiliser codes as
follows.

\begin{definition}
  Let $S$ be a isotropic subset of $\mathbb{F}_p^{n}\times
  \mathbb{F}_p^n$ associated to a linear cyclic stabiliser code
  $\mathcal{C}$. The \emph{BCH distance of $\mathcal{C}$} is the BCH
  distance of the generator polynomial of the centraliser
  $\centraliser{S}$.
\end{definition}

We have the following theorem which follows from
Theorem~\ref{thm:isotropic-main-properties}.

\begin{theorem}
  Let $\mathcal{C}$ is be any linear cyclic stabiliser code of BCH
  distance $d$. Then it is $d$-pure and hence has distance at least
  $d$.
\end{theorem}

\section{Linear cyclic codes of length dividing $p^{t}+1$}

In this section, we study linear cyclic stabiliser codes over
$\mathbb{F}_p$ whose length divides $p^{t} +1$. The main motivation to restrict
our attention to lengths of this form is captured in the following proposition.

\begin{proposition}\label{prop:inverse-frob}
If the integer $n$ divides $p^t+1$, for some positive integer $t$ then $X^{-1}$
in the cyclotomic
ring $\mathbb{F}_p[X]/X^{p^t+1} -1$ is $X^{p^{t}}$. Therefore, for
every polynomial $g(X)$ over any extension of $\mathbb{F}_p$ we have $g(X^{-1})$ is
$g(X)^{p^t}$.
\end{proposition}

The above-mentioned property simplifies the isotropy condition for polynomials
considerably and allows us to completely characterise all linear
cyclic codes of such lengths.

Let $\mathbb{F}_p(\eta)/\mathbb{F}_p$ be an extension of degree $d$. 
When dealing with cyclic
quantum codes of length $n$, we use $\R$ to denote the cyclotomic ring
$\mathbb{F}_p[X]/X^n -1$. The extension ring $\Ra$ is then the
cyclotomic ring $\mathbb{F}_{p}(\eta)[X]/X^n -1$. Linear codes are
associated with quadratic extension and identification of the pair of
vectors $(\Va,\Vb)$ with $\Va+\eta \Vb$ maps its isotropic set to an ideal
of $\Ra$.


\begin{lemma}\label{lem:cyclic-linear}
  Let $S$ be the isotropic ideal associated to a linear cyclic
  stabiliser code over $\mathbb{F}_p$ of length dividing $p^t+1$. Then $S$ is
  uniquely cyclic.
\end{lemma}
\begin{IEEEproof}
Let $\mathbb{F}_p(\eta)/\mathbb{F}_p$ be the quadratic extension such
that $S$ is an ideal of the cyclotomic ring $\mathbb{F}_p(\eta)[X]/(X^n-1)$. Let$X^2+c_1 X+c+0$ be the minimal polynomial of $\eta$ over $\mathbb{F}_p$.
 
  Recall that the projection of $S$ onto the first $n$ co-ordinates
  forms a classical cyclic code over $\mathbb{F}_p$ and hence is
  generated by a factor $g(X)$ of $X^n-1$.  Suppose there exist two
  distinct elements $(\mathbf{g},\mathbf{f})$ and
  $(\mathbf{g},\mathbf{f'})$ in $S$.  Let
  $\mathbf{h}=\mathbf{f}-\mathbf{f'}$ so that $(0,\mathbf{h})$ is also
  in $S$. We prove that the polynomial $h(X) = 0$ modulo $X^n -1$. By
  the Chinese remaindering theorem, it is sufficient to prove
  separately that all roots of $\frac{X^n-1}{g}$ and $g$ are roots of
  $h$.

  From Proposition~\ref{prop:inverse-frob} we get, $g(X^{-1}) =
  g^{p^t}(X)$. Applying Proposition~\ref{prop:isotropy-polynomials} to
  the elements $(0,\mathbf{h})$ and $(\mathbf{g},\mathbf{f})$ we have
  $g^{p^t}h=0 \mod X^n-1$. Since $g$ is invertible modulo $\frac{X^n -
    1}{g}$, every root of $\frac{X^n - 1}{g}$ should also be a root of
  the polynomial $h(X)$.

  We now show that every root of $g$ is also a root of $h$. Since
  $\eta h$ belongs to $S$, if the code is linear then $\eta^2 h=-c_0 h
  - \eta c_1 h$ must also belong to $S$ where $\eta$ is a root of the
  quadratic polynomial $X^2+c_1 X+c_0$.  Any element in $S$ is of the
  form $ag+\eta b$, where $a(X)$ and $b(X)$ are polynomials in
  $\mathbb{F}_p[X]/X^n-1$. Hence, $-c_0h = ag$ and every root of $g$ is
 also a root of $h$.
\end{IEEEproof}

Consider the Frobenius automorphism $\sigma$ on a degree $d$ extension
$\mathbb{F}_p(\eta)/\mathbb{F}_p$ which maps any element $\alpha$ in
 $\mathbb{F}_p(\eta)$ to $\alpha^p$. This can be naturally
 extended to
polynomials over $\mathbb{F}_p(\eta)$ and therefore on $\Ra$ as
follows: For a polynomial $a(X) = a_0 + \ldots + a_n X^n$ where $a_i$
are in $\mathbb{F}_p(\eta)$, $\sigma(a)$ is defined as $\sigma(a_0) +
\ldots + \sigma(a_n) X^n$. We call this the \emph{Frobenius
  involution}.

Constructing linear cyclic codes correspond to constructing generators
for the associated isotropic ideal. We make use of the following
Galois theoretic lemma to characterise such generators. 

\begin{lemma}\label{lem:degree} 
  Let the integer $n$ divide $p^t+1$ for some positive integer $t$. 
\begin{enumerate}
\item Any irreducible factor of $X^{n} -1$ over
  $\mathbb{F}_p$ other than the factors $X -1$ or $X+1$ has even degree.
\item Let $f(X)$ be any irreducible factor of $X^n-1$ over $\mathbb{F}_p$ whose degree is divisible by $d$ for some positive integer $d$. Over the extension field $\mathbb{F}_{p^d}=\mathbb{F}_p(\eta)$, $f(X)$ splits into $d$ irreducible factors $f_0(X,\eta), \ldots, f_{d-1}(X,\eta)$  such that $f_i = \sigma^i(f_0)$.
\end{enumerate}
\end{lemma}
\begin{IEEEproof}[Proof of part 1]
  Consider any irreducible factor $f(X)$ of $X^n -1$ over
  $\mathbb{F}_p$ other than $X-1$ or $X+1$. Let $k$ be the degree of
  $f(X)$. Then, the splitting field of $f(X)$ over $\mathbb{F}_p$ is
  $\mathbb{F}_{p^k}$. Consider any root $\beta$ of $f(X)$ in
  $\mathbb{F}_{p^k}$. The Frobenius automorphism $\sigma^{t}$ is a
  field automorphism of $\mathbb{F}_{p^k}$ and $\sigma^{t}(\beta) =
  \beta^{p^{t}}$. Notice that $\beta$ is an $n$-th root of unity and
  $n$ divides $p^t+1$. Hence $\sigma^{t}(\beta) = \beta^{-1}$ and
  $f(\beta^{-1}) = \sigma^{t} (f(\beta)) = 0$.  Since $f(X)$ is
  neither $X - 1$ nor $X+1$, we have $\beta \neq \pm 1$ and hence
  $\beta \neq \beta^{-1}$. As a result, the roots of $f(X)$ comes in
  pairs; for every root $\beta$ its inverse $\beta^{-1}$ is also a
  root. Hence, the degree $k$ of $f(X)$ should be an even number.
\end{IEEEproof}

\begin{IEEEproof}[Proof of part 2]
  Consider any irreducible factor $f(X)$ of $X^n -1$ of degree $k=dm$
  for some positive integer $m$. Its splitting field $\mathbb{F}_{p^k}
  = \mathbb{F}_{p^{dm}}$ therefore, contains $\mathbb{F}_{p^d}$. Any
  irreducible factor of $f(X)$ over $\mathbb{F}_{p^d}$ should be of
  degree equal to the degree of the extension
  $\mathbb{F}_{p^k}/\mathbb{F}_{p^d}$ which is $m$. The Frobenius
  $\sigma$ being a field automorphism of $\mathbb{F}_{p^d}$, should
  map these factors to each other. Further, order of $\sigma$ in
  $\mathbb{F}_{p^d}$ is $d$. Thus $f(X) = f_0(X,\eta)\cdot \ldots
  \cdot \sigma^{d-1}(f_0(X,\eta))$ over $\mathbb{F}_{p^2}$.
\end{IEEEproof}

Consider the extension field $\mathbb{F}(\eta)=\mathbb{F}_{p^2}$ and
let $S$ be any ideal of $\Ra$. The following theorem gives a
necessary condition for it to be isotropic and hence give a linear
cyclic code. 

\begin{theorem}\label{thm:linear-necessary-condition}
  Let $\mathbb{F}_p(\eta)$ be a quadratic extension of $\mathbb{F}_p$.
 Let $n$ divide $p^t +1$ and $S$ be an isotropic ideal of
  $\mathbb{F}_p(\eta)[X]/X^n-1$. Then $t$ is even and the ideal $S$ is
  generated by the product polynomial $g(X) \cdot h(X,\eta)$ where
  $g(X)$ and $h(X,\eta)$ are two coprime factors of $X^n -1$
  satisfying the following condition.
  \begin{enumerate}
  \item \label{prop-g-polynomial} $g(X)$ is any factor of $X^n - 1$
    over $\mathbb{F}_p$ which contains both $X-1$ and $X+1$ as
    factors.
  \item \label{prop-h-polynomial} $h(X,\eta)$ is any factor of
    $\frac{X^n -1}{g}$ over $\mathbb{F}_{p^2}$, such that for any
    irreducible factor $r(X,\eta)$ of $\frac{X^n -1}{g}$ over
    $\mathbb{F}_{p^2}$, $r(X,\eta)$ divides $h(X,\eta)$ if and only if
    $\sigma(r) = r(X,\eta')$ \emph{does not}.
  \end{enumerate}
\end{theorem}
\begin{IEEEproof}
  From Lemma~\ref{lem:cyclic-linear} it follows that $S$ is uniquely
  cyclic. Let $(g,f)$ be a generating pair for $S$ where $g(X)$ and
  $f(X)$ are polynomials over $\mathbb{F}_p$. Then the polynomial
  $g(X) + \eta f(X)$ is an element of the ideal $S$. It follows from
  the linearity of $S$ that the polynomial $\eta(g+\eta f)$ is also in
  $S$. However, the set $S$ is uniquely cyclic. Using
  Proposition~\ref{prop:uniquely-cyclic}, there is a polynomial $a(X)$
  in $\mathbb{F}_p[X]$ such that
  \begin{equation}\label{eqn:linearity-condition}
    \eta(g+\eta f) = a(g+\eta f) \mod X^n-1
  \end{equation}

  Let $c(X)=X^2+c_1 X + c_0$ be the minimal polynomial of $\eta$ over
 $\mathbb{F}_p$
  where $c_0$ and $c_1$ are elements of $\mathbb{F}_p$. Comparing the
  coefficients of $\eta$ in Equation~\ref{eqn:linearity-condition}, we
  have
  \begin{eqnarray}
    f &=& - \frac{a}{c_0}  g \mod X^n - 1 \textrm{ and} \label{eqn:f}\\
    c(a(X)) &=& 0 \mod \frac{X^n-1}{g}.\label{eqn-c-a}
  \end{eqnarray}

  When $n$ divides $p^t+1$, for any polynomial $\gamma(X)$ in the cyclotomic
  ring $\mathbb{F}_p[X]/X^n-1$, $\gamma(X^{-1})$ is just
  $\gamma^{p^t}(X)$.  Since $S$ is isotropic, from
  Equation~\ref{eqn:f} and Proposition~\ref{prop:isotropy-polynomials}
  its follows that $g^{p^t+1} a^{p^t}=g^{p^t+1} a \mod X^n -1$. The polynomial
  $g(X)$ is invertible modulo $\frac{X^n-1}{g}$. As a result we have,
  \begin{equation}\label{eqn:iso-restated}
   a^{p^t} = a \mod  \frac{X^n -1}{g}
  \end{equation}

  Let $r(X)$ be any irreducible factor of $\frac{X^n-1}{g}$ over
  $\mathbb{F}_p$ and $\mathbb{K}$ be the extension field
  $\mathbb{F}_p[X]/r(X)$. From Equation~\ref{eqn-c-a} we have, $a \mod
  r$ is a root of the polynomial $c(Y)$ over the extension field
  $\mathbb{K}$. If possible, let $t$ be an odd integer $2m+1$.  Since
  $c(Y)$ divides $Y^{p^{2m}}-Y$, $a^{p^{2m}}=a \mod r$. Using
  Equation~\ref{eqn:iso-restated} we get $a^p = a$ modulo $r$ and
  hence $a \mod r$ is an element of the sub field $\mathbb{F}_p$ of
  $\mathbb{K}$. However, this is a contradiction, since the polynomial
  $c$ is irreducible over $\mathbb{F}_p$. Therefore, $t$ must be even.

  Recall that $a \mod r$ is a root of the polynomial $c(Y)$ over the
  extension field $\mathbb{K}$. This implies that the extension field
  $\mathbb{K}$ contains $\mathbb{F}_p[Y]/c(Y) = \mathbb{F}_{p^2}$.
  Therefore, degree of $r$ must be even and $g$ must have as factors
  all the odd degree irreducible factors of $X^n -1$ over
  $\mathbb{F}_p$. By Lemma~\ref{lem:degree}, these odd degree factors
  are just $X-1$ and $X+1$. Thus $g$ satisfy
  property~\ref{prop-g-polynomial} of the theorem.

  Consider the polynomial $h(X,\eta) =
  \mathrm{gcd}\left(\frac{X^n-1}{g},1 - \frac{\eta}{c_0}
  a\right)$. Clearly $h$ is coprime to $g(X)$. We claim that $g \cdot
  h$ generates the ideal $S$.  To see this, notice that $S$ as a
  subspace of $\mathbb{F}_p^n \times \mathbb{F}_p^n$ is uniquely
  cyclic and is generated by the pair $(g,f)$, where $f =
  -\frac{a}{c_0} g$ modulo $X^n -1$ (using Equation~\ref{eqn:f}).
  Therefore, $S$ as an ideal is also generated by the polynomial
  $\mathrm{gcd}\left(X^n -1, g\left(1- \frac{\eta}{c_0}
  a\right)\right)$ which is the product $g \cdot h$. We claim that
  polynomial $h$ thus constructed satisfies the properties mentioned
  in the theorem. Any irreducible factor $r(X)$ of $\frac{X^n -1}{g}$
  over the field $\mathbb{F}_p$ is of even degree and hence factorises
  as $r_1(X,\eta)\sigma(r_1(X,\eta))$ over
  $\mathbb{F}_p(\eta)$. Recall that $a \mod r$ is a root of $c(Y)$.
  As a result, $a \mod r_1$ is either $\eta$ or $\eta'$. Now, $r_1$
  divides $h$ if and only if $1- \frac{\eta}{c_0} a$ modulo $r_1$ is
  zero. Therefore, $r_1$ divides $h$ if and only if $a = \eta' \mod
  r_1$.  The polynomial $a$ has coefficients in $\mathbb{F}_p$ and
  hence $a = \sigma(a)$. As a result by the third property of
  Proposition~\ref{prop:frobenius-involution}, $a = \eta' \mod r_1$ if
  and only if $a = \sigma(\eta') = \eta \mod \sigma(r_1)$.  For each
  pair $r_1$ and $\sigma(r_1)$, exactly one of them divide $h$
  depending one whether $a$ is $\eta$ or $\eta'$ modulo $r_1$.  This
  proves the theorem.
\end{IEEEproof}

A corollary of the above theorem is the following impossibility
result.

\begin{corollary}\label{cor:nonexistance-power-odd}
  Let $n$ be any integer that divides $p^t+1$, where $t$ is odd. Then there
  does not exist any linear cyclic stabiliser codes of length $n$ over
  $\mathbb{F}_p$.
\end{corollary}

For example, $9,11,19,27,33,43,57,59,67,81,83,99$ are the numbers less
then hundred that divide $2^t+1$ for some odd $t$. Hence there is no binary
linear cyclic code of such lengths.

The next theorem shows that the conditions in
Theorem~\ref{thm:linear-necessary-condition} are also sufficient to
construct isotropic ideals of $\Ra$. This gives us a way of
constructing linear cyclic stabiliser of length dividing $p^{2m} +1$. This theorem directly follows from a more generalised construction given in Theorem~\ref{thm:gen-const} and Theorem~\ref{thm:connection}.

\begin{theorem}\label{thm:main-construction}
  Let $n$ divide $p^{2m} +1$ and $\mathbb{F}_p(\eta)$ be a quadratic
extension of $\mathbb{F}_p$. Let $g(X)$ and $h(X,\eta)$ be factors of
  $X^n -1$ satisfying the properties~\ref{prop-g-polynomial} and
  \ref{prop-h-polynomial} of
  Theorem~\ref{thm:linear-necessary-condition}.  Then the ideal $S$ of
  $\mathbb{F}_p(\eta)[X]/X^n-1$ generated by the product $g \cdot h$
 is isotropic as a subset
  of $\mathbb{F}_p^n \times \mathbb{F}_p^n$ and the associated
  stabiliser code is linear and cyclic.
\end{theorem}

In the rest of the article, we refer to
cyclic stabiliser codes whose length divide $p^{t}+1$ as $t$-Frobenius
codes. For linear $2m$-Frobenius codes, we call the factorisation $g(X) \cdot
h(X,\eta)$ characterised above as the \emph{canonical factorisation}
associated to the code.

\begin{theorem}
  Let $\mathcal{C}$ be a linear $2m$-Frobenius code over
  $\mathbb{F}_p$ with canonical factorisation $g
  \cdot h$. The stabiliser dimension of the code
  $\mathcal{C}$ is $\mathrm{deg}(g)$. The centraliser $\centraliser{S}$
 of $S$ is the ideal generated by $h(X,\eta)$ and hence the BCH distance of
 $\mathcal{C}$ is BCH distance of $h$. 
\end{theorem}
Again the proof follows from the more general theorem \ref{thm:gen-prop} and \ref{thm:connection}.

\section{Generalisation to nonlinear codes}
We have already shown that if $n$ divides $p^t+1$ for some odd integer $t$
 then no linear code of length $n$ exists. In this section we show how to
construct nonlinear codes of such length. The construction is a
generalisation of Theorem~\ref{thm:main-construction}. The major difference
is that the extension of $\mathbb{F}_p$ is no longer restricted
to be quadratic. 

\begin{theorem}\label{thm:gen-const}
  Let $n$ divide $p^{dm}+1$ and $\mathbb{F}_p(\eta)$ be a degree $d$
  extension of $\mathbb{F}_p$. Let $g(X)$ and $h(X,\eta)$ be co-prime factors
  of $X^n-1$ satisfying the following properties.
  \begin{enumerate}
  \item $g(X)$ is any factor of $X^n-1$ over $\mathbb{F}_p$ which
    contains all the the irreducible factor of $X^n-1$ over
    $\mathbb{F}_p$ whose degree is not divisible by $d$.
  \item $h(X,\eta)$ is any factor of $\frac{X^n-1}{g}$ over
    $\mathbb{F}_p(\eta)$ such that for any irreducible factor
    $r(X,\eta)$ of $\frac{X^n-1}{g}$ over $\mathbb{F}_p(\eta)$,
    $r(X,\eta)$ divides $h(X,\eta)$ if and only if none of the factors
    $\sigma(r),\ldots,\sigma^{d-1}(r)$ divide $h$
    i.e. $\frac{X^n-1}{g(X)}=\prod_{i=0}^{d-1}\sigma^i(h)$.
  \end{enumerate}
  Fix any nonzero $\alpha$ in $\mathbb{F}_p$ and let $a(X,\eta)$ be
  the polynomial, uniquely defined by Chinese remaindering, as
  follows.

  \[
  a = \left\{\begin{array}{ll}
  1 & \mod g \\
  \sigma^i(\alpha\eta) & \mod \sigma^i(h) \mbox{ for all } 0\leq i <d \\
  \end{array}\right.
  \]
  Then $a(X,\eta)$ is a polynomial in $\mathbb{F}_p[X]$ and the
  uniquely cyclic subspace generated by $(g,ag)$ is isotropic.
\end{theorem}
The proof of this theorem, involves verifying certain equations modulo
$X^n - 1$. Almost always we do this in by verifying the said equation
separately modulo $g$ and each irreducible factor $r(X,\eta)$ of
$\frac{X^n -1}{g}$. Then by Chinese remaindering, we have the said
equation modulo $X^n -1$. We call this the \emph{Chinese remainder
  verification}.

Let us call the polynomial $\frac{X^n-1}{g(X)}$ as $f(X)$. From the
definition of the factor $h(X,\eta)$ it follows that over
$\mathbb{F}_p(\eta)$, the polynomial $f(X)$ splits as $h\cdot
\sigma(h) \cdots \sigma^{d-1}(h)$.

\begin{claim}\label{claim:ainfp}
    The polynomial $a(X,\eta)$ is a polynomial in $\mathbb{F}_p[X]$
\end{claim}

\begin{IEEEproof} 
  It is sufficient to prove that $\sigma(a) = a$. Notice that since
  $a=1 \mod g$, $\sigma(a) = a \mod g$. Using
  Proposition~\ref{prop:frobenius-involution}, it is sufficient to
  show that $\sigma(a) = a \mod f$. Since $\sigma^d(h)=h$ and $a =
  \sigma^{i}(\alpha\eta) \mod \sigma^{i}(h)$ for all $i$, applying
  $\sigma$ on both side we get $\sigma(a)=\sigma^i(\alpha\eta) \mod
  \sigma^i(h)$ for all $i$. By Chinese remainder verification the
  claim follows.
\end{IEEEproof}

\begin{claim}\label{clm:a-power-n}
$a(X)=a(X^{-1}) \mod f(X)$
\end{claim}
\begin{IEEEproof}
  From Proposition~\ref{prop:inverse-frob} we have,
  $a(X^{-1})=\sigma^t(a(X)) \mod X^n-1$. We know, $a =
  \sigma^i(\alpha\eta) \mod \sigma^i(h)$. Since
  $\sigma^t(\alpha\eta)=\alpha\eta$, applying $\sigma^t$ on $a$ we get
  $\sigma^t(a) = \sigma^i(\alpha\eta) \mod \sigma^i(h)$.  Hence by
  Chinese remaindering, $\sigma^t(a)=a \mod f$.
\end{IEEEproof}

\begin{claim}
The uniquely cyclic subspace generated by the pair $(g,ag)$ is isotropic.
\end{claim}
\begin{IEEEproof}
Using Proposition~\ref{prop:isotropy-uniquely-cyclic}, it is is sufficient
  to prove that
  \begin{equation}\label{eqn-isotropy-a-g}
    g(X) a(X^{-1}) g(X^{-1}) - g(X^{-1}) a(X) g(X) = 0 \mod X^n -1.
  \end{equation}
  Clearly Equation~\ref{eqn-isotropy-a-g} holds modulo $g(X)$ as both
  the terms are divisible by $g$. From Proposition~\ref{prop:inverse-frob}
  and Equation~\ref{clm:a-power-n} we have, $a(X^{-1}) =  a$
  modulo $\frac{X^n -1}{g}$. We
  then apply Chinese remaindering and conclude that the subspace is isotropic.
\end{IEEEproof}

The following theorem shows that the linear codes obtained from Theorem~\ref{thm:main-construction} are indeed a subclass of the codes
generated from Theorem~\ref{thm:gen-const}
\begin{theorem}\label{thm:connection}
Let $c(X)=X^2+c_1X+c_0$ be an irreducible polynomial over $\mathbb{F}_p$ and $\eta,\eta'$ be roots of $c(X)$. Fix $d=2$, $\mathbb{F}_p(\eta')/\mathbb{F}_p$ to be the extension and $\alpha=-c_0^{-1}$ in Theorem~\ref{thm:gen-const} and let $S$ be the corresponding isotropic subspace. Then the image
of $S$ under the map $(u,v)\mapsto u+\eta v$ is an ideal of the cyclotomic
ring $\mathbb{F}_p(\eta)[X]/(X^n-1)$ and its generator is given by
the polynomial $g(X)h(X,\eta)$ where $g,h$ satisfies the properties in Theorem~\ref{thm:linear-necessary-condition}. Moreover the centraliser $\centraliser{S}$ also maps to an the ideal generated by $h$.
\end{theorem}
\begin{IEEEproof}
We know from Theorem~\ref{thm:gen-const} that the polynomial $a(X,\eta)$ defined by the following
\begin{equation}\label{eqn:a-def}
 a = \sigma^i(-c_0^{-1}\eta') \mod \sigma^i(h) \mbox{ where } i \in \{0,1\}
\end{equation}
belongs to the ring $\mathbb{F}_p[X]/g(X)$ and the uniquely cyclic subspace $S$
generated by the pair $(g,ag)$ is isotropic.

To prove that $S$ maps to an ideal it is sufficient to show that for any
element $(ug,uag)$ in $S$ there exists another element $(vg,vag)$ in $S$
such that $\eta(ug+\eta uag)=v+\eta vag \mod X^n-1$. We claim that 
we can always choose  $v=c_0 au$ to satisfy this condition.
\begin{claim}
$\eta (g+\eta ag) =c_0 a(g+\eta ag) \mod X^n-1$
\end{claim} 
\begin{proof}
The cyclotomic polynomial $X^n-1$ is product of $g$,$h$ and $\sigma(h)$.
Using equation~\ref{eqn:a-def} and the fact that $c_0=\eta\eta'$, $-\eta^2=c_1\eta+c_0$ it is straight-forward to verify the claim separately modulo $g(X)$, $h$ and $\sigma(h)$. Then by Chinese remaindering conclude that it is true modulo $X^n-1$.
\end{proof}

Equation~\ref{eqn:a-def} implies that for any irreducible factor $r(X,\eta)$ of $\frac{X^n-1}{g}$ over $\mathbb{F}_p(\eta)$
\begin{equation}\label{eqn:a-alt} a \mod r = \left\{\begin{array}{ll}
-\eta^{-1}  & \mbox{ if } r | h \\
-c_0^{-1}\eta & \mbox{ if } r | \sigma(h) \\ \end{array}\right.
\end{equation}
Now $\tilde{g}(X,\eta)=(g+\eta ag)$ is a generator of $S$ as an ideal and equation~\ref{eqn:a-alt} imply that if $r|h$ then $\tilde{g} \mod r$ is zero and if $r|\sigma(h)$ then $\tilde{g} \mod r$ is $\frac{c_1}{c_0}\eta +1$ which is nonzero. Therefore $\mathrm{gcd}(X^n-1,\tilde{g})=gh$ and $gh$ is also a generator of $S$.

Let $I$ be the ideal generated by $1+\eta a$. Since $\mathrm{gcd}(X^n-1,1+\eta a)=h$, the ideal $I$ is also generated by $h$. We know that $a(X)=a(X^{-1}) \mod \frac{X^n-1}{g}$ (see proof of Claim~\ref{clm:a-power-n}). By Proposition~\ref{prop:isotropy-uniquely-cyclic} it can be verified that any element of the form $(u,ua)$ belongs to the centraliser. Since $\centraliser{S}$ itself is an ideal, $I$ is a subideal of $\centraliser{S}$. To show that $I$ is actually $\centraliser{S}$ we show that they have same cardinality. The cardinality of $I$ is $({p^2})^{n-deg(h)}$ which is equal to $p^{n+deg(g)}$, since $\deg(g)+2 \deg(h)=n$. On the other hand cardinality of $S$ is $p^{n-deg(g)}$. Hence by Theorem~\ref{thm:isotropic-main-properties} cardinality of $\centraliser{S}$ is $p^{n+deg(g)}$.
\end{IEEEproof}

As before, we call $g\cdot h$ as the canonical factorisation associated with the above mentioned $t$-Frobenius codes. We also
call the BCH distance of $h$ to be the BCH distance of $\mathcal{C}$.

\begin{theorem}\label{thm:gen-prop}
Let $g(X)\cdot h(X,\eta)$ be the canonical factorisation associated with
a $t$-Frobenius code $\mathcal{C}$ as in Theorem~\ref{thm:gen-const}. The
stabiliser dimension of $\mathcal{C}$ is $\deg(g)$. If the BCH distance of $h$ is $\delta$ then $\mathcal{C}$ is $\delta$-pure and
hence has distance at least $\delta$. 
\end{theorem}
\begin{proof}
Let $a(X)$ be the polynomial corresponding to $\mathcal{C}$ such that the
isotropic subspace $S$ of $\mathcal{C}$ is generated by the pair $(g,ag)$ as in
Theorem~\ref{thm:gen-const}. Since any element in $S$ is of the form $(ug,uag)$, the number of distinct values $u$ can take is the cardinality of the ring
$\mathbb{F}_p[X]/g(X)$ which is $p^{n-\deg(g)}$. Hence by
 Theorem~\ref{thm:isotropic-main-properties} we conclude that the stabiliser dimension of $\mathcal{C}$ is $\deg(g)$

To prove the lower bound on distance we first need the following result

\begin{claim}\label{clm:cent}
  Any element in the centraliser $\centraliser{S}$ is of the form
  $\left(u,au+v\frac{X^n-1}{g}\right)$ for some polynomials $u(X)$ and
  $v(X)$ over $\mathbb{F}_p$ such that $g(X)$ and $v(X)$ are coprime.
\end{claim}
\begin{IEEEproof}
  Let $A$ be the set of all pairs $\left(u,au + v \frac{X^n
    -1}{g}\right)$ where $u(X)$ and $v(X)$ are polynomials over
  $\mathbb{F}_p$ such that $v(X)$ is coprime to $g(X)$. It follows
  from Proposition~\ref{prop:isotropy-uniquely-cyclic} that the set
  $A$ is contained in $\centraliser{S}$. However, the cardinality of
  $A$ is the product of the cardinalities of the rings
  $\mathbb{F}_p[X]/(X^n-1)$ and $\mathbb{F}_p[X]/g(X)$ which is
  $p^{n+\deg(g)}$. By Theorem~\ref{thm:isotropic-main-properties},
  cardinality of $\centraliser{S}$ itself is $p^{n+\deg(g)}$. Hence
  $\centraliser{S}$ is equal to the set $A$.
\end{IEEEproof}

Notice that the joint weight of a pair $(u'(X),v'(X))$ is equal to the
weight of $\alpha\eta u' - v'$ as a polynomial over
$\mathbb{F}_p(\eta)$.  We know, for any factor $r(X,\eta)$ of $X^n-1$,
 weight of any polynomial over $\mathbb{F}_p(\eta)$ which is a multiple of $r$,
 is at least the BCH
distance of $r$. Therefore, to prove that $\mathcal{C}$ is $\delta$-pure it
is sufficient to show that for any element $(u',v')$ in the
centraliser $\centraliser{S}$ the polynomial $h(X,\eta)$ is a factor
of $\alpha\eta u' - v'$. By Claim~\ref{clm:cent} we know that there
exists a polynomials $v$ such that $v'=au'+v\frac{X^n-1}{g}$. Since
$h$ divide $\frac{X^n-1}{g}$ and $a = \alpha\eta \mod h$, it follows
that $\alpha\eta u' - v' \mod h = 0$.
\end{proof}

As a demonstration of our construction we list
(Table~\ref{tab:explicit-examples-short}) some explicit examples of
codes where the characteristic $p$ of the underlying finite field is
2. The distance given in this table is the BCH distance.  The actual
distance can be larger. Canonical factors and their roots 
 are given in the appendix. We have both linear and non-linear codes for parameters with dagger whereas star denotes only nonlinear codes.
\begin{table}[h]
\begin{center}
\begin{tabular}{|l|l|}
\hline
Length & Parameters\\
\hline
5 & [[5,1,3]] \\ 
\hline
9 & [[9,3,3]]* \\
\hline
13 & [[13,1,5]] \\ 
\hline
17 & [[17,1,7]] , [[17,9,3]] \\ 
\hline
19 & [[19,1,3]]* \\
\hline
25 & [[25,1,4]] , [[25,5,3]] \\ 
\hline
27 & [[27,21,2]]*, [[27,9,3]]* \\
\hline
29 & [[29,1,5]] \\ 
\hline
37 & [[37,1,5]] \\ 
\hline
41 & [[41,1,7]] , [[41,21,4]] \\ 
\hline
53 & [[53,1,7]] \\ 
\hline
57 & [[57,21,5]]*, [[57,39,3]]* \\
\hline
61 & [[61,1,7]] \\ 
\hline
\multirow{2}{*}{65} & [[65,5,13]]*, [[65,13,8]] , [[65,17,9]], [[65,17,11]]* ,\\
	& [[65,29,7]]$^\dagger$ , [[65,41,5]]$^\dagger$ , [[65,53,3]]$^\dagger$ \\ 
\hline
67 & [[67,1,7]]* \\
\hline
81 & [[81,21,4]]*, [[81,75,2]]*\\
\hline
97 & [[97,1,9]] , [[97,49,5]] \\ 
\hline
99 & [[99,69,3]]* \\
\hline
\end{tabular}
\vspace{-1mm}
\end{center}
\caption{Explicit examples of Frobenius codes over $\mathbb{F}_2$}
\label{tab:explicit-examples-short}
\vspace{-1cm}
\end{table}
\section{Decoding}

Let $\mathcal{C}$ be a $t$-Frobenius code based on a degree $d$ extension $\mathbb{F}_p(\eta)$ as in Theorem~\ref{thm:gen-const}. Let the code $\mathcal{C}$ have length $n$ and  BCH distance $\delta=2\tau +1$. Much like in the classical case, we show that there is an poly$(n)$ time quantum algorithm to correct any quantum error of weight at most $\tau$. We use two key algorithms: (1) Kitaev's
phase estimation~\cite[5.2]{nielsen:and:chuang} algorithm and (2) The
Berlekamp decoding algorithm~\cite[p-98,6.7]{vLin} for classical BCH
codes.

\begin{theorem}[Berlekamp]\label{thm:berlekamp-recast}
  Let $h(X)$ be a factor of $X^n - 1$ of BCH distance $\delta =2 \tau +1$ over
  a finite field $\mathbb{F}_q$, $q$ and $n$ coprime. Let $e(X)$ be
  any polynomial of weight at most $\tau$ over $\mathbb{F}_q$. Given a
  polynomial $r(X) = e(X) \mod h(X)$, there is a polynomial time
  algorithm to find $e(X)$.
\end{theorem}

Let the canonical factorisation of the $\mathcal{C}$ be $g\cdot h$ so that its isotropic subspace is generated by the pair $(g,ag)$ where $a = \sigma^i(\alpha\eta) \mod \sigma^i(h)$. Assume that we
transmitted a quantum message $\ket{\varphi} \in \mathcal{C}$ over the
quantum channel and received the corrupted state $\ket{\psi} =
U_\mathbf{u} V_\mathbf{v} \ket{\varphi}$, where the vectors
$\mathbf{u}$ and $\mathbf{v}$ are unknown but fixed for the rest of
the section. We show that using quantum phase finding we can recover the polynomial $\alpha\eta u(X^{-1})-v(X^{-1}) \mod h$ without disturbing $\ket{\psi}$. Provided the joint weight w$(\mathbf{u},\mathbf{v})\leq \tau$ we can now find $\mathbf{u}$ and $\mathbf{v}$ using Berlekamp algorithm. The sent message is recovered by applying the
inverse map $V_\mathbf{v}^\dag U_\mathbf{u}^\dag$ on $\ket{\psi}$. Hence we have the following theorem about decoding linear $t$-Frobenius codes.

\begin{theorem}\label{thm:main-decoding-theorem}
  Let $\mathcal{C}$ be a $t$-Frobenius code, as in Theorem~\ref{thm:gen-const}, of length
  $n$ and BCH distance $\delta = 2\tau+1$. There is quantum
  algorithm that takes time polynomial in $n$ to correct errors of
  weight at most $\tau$.
\end{theorem}
\begin{IEEEproof}
Let $S$ be the isotropic subspace corresponding to $\mathcal{C}$. We need the following lemma to prove the theorem.

\begin{lemma}\label{lem:compute-poly-iso}
  Let $\ket{\varphi}$ be a codeword in $\mathcal{C}$ and $\ket{\psi} =
U_\mathbf{u} V_\mathbf{v} \ket{\varphi}$, where $\mathbf{u}$ and $\mathbf{v}$ are any vectors in $\mathbb{F}_p^n$. There is an efficient quantum algorithm which, given the state $\ket{\psi}$ and any element $(\Va,\Vb)$ in $S$, computes the polynomial $b(X) u(X^{-1}) -  a(X)v(X^{-1}) \mod X^n-1$ without destroying $\ket{\psi}$.
\end{lemma}
\begin{IEEEproof}
  Recall that for every $(\mathbf{a},\mathbf{b})$ in $S$ and
  $\ket{\varphi}$ in $\mathcal{C}$, the operator
  $W_{\mathbf{a},\mathbf{b}} =
  \omega^{\transpose{\mathbf{a}}\mathbf{b}} U_\mathbf{a} V_\mathbf{b}$
  stabilises $\ket{\varphi}$ (Theorem~\ref{thm:iso-stab-connection}).
  Suppose the received vector is $\ket{\psi} = U_{\mathbf{u}}
  V_{\mathbf{v}} \ket{\varphi}$. It is easy to verify that the vector
  $\ket{\psi}$ is an eigen vector of the operator
  $W_{\mathbf{a},\mathbf{b}}$ and the associated eigenvalue is
  $\zeta^{\transpose{\mathbf{b}}\mathbf{u} -
    \transpose{\mathbf{a}}\mathbf{v}}$ where $\zeta$ is the primitive
  $p$-th root of unity. One can recover this phase without disturbing
  $\ket{\psi}$ using quantum phase finding. Repeating the algorithm
  with $(N^k\mathbf{a},N^k\mathbf{b})$, all the inner products
  $\transpose{\mathbf{b}}N^k \mathbf{u} - \transpose{\mathbf{a}}N^k
  \mathbf{v}$ can be recovered. These are precisely the coefficients
  the polynomial $b(X) u(X^{-1}) - b(X)v(X^{-1})$ modulo $X^n
  -1$. Hence proved.
\end{IEEEproof}

Let $\mathcal{C}$ be based on a degree $d$ extension $\mathbb{F}_p(\eta)$ as in Theorem~\ref{thm:gen-const}. Let the canonical factorisation associated with $\mathcal{C}$ be $g(X)\cdot h(X,\eta)$. The isotropic subspace $S$ of $\mathcal{C}$ is generated by the pair $(g,ag)$ where $a = \sigma^i(\alpha\eta) \mod \sigma^i(h)$. Using Lemma~\ref{lem:compute-poly-iso} we can compute the polynomial
 $e'(X)=a(X)g(X)u(X^{-1})-g(X)v(X^{-1}) \mod X^n-1$. The factor $g(X)$ has an inverse $g^{-1}(X)$ in the ring $\mathbb{F}_p[X]/\frac{X^n-1}{g(X)}$ i.e. $gg^{-1}=1 \mod \frac{X^n-1}{g}$. Since $h$ is a factor of $\frac{X^n-1}{g}$ and $a=\alpha\eta \mod h$, multiplying $e'$ by $g^{-1}$ and taking modulo $h$ we get the polynomial $e(X,\eta)=\alpha\eta u(X^{-1})-v(X^{-1}) \mod h(X,\eta)$.

Remember that for any vector $\mathbf{u}=(u_0,\ldots,u_{n-1})$ in
$\mathbb{F}_p^n$ we have $u(X^{-1})=u'(X)$ where
$\mathbf{u'}=(u_0,u_{n-1},\ldots,u_1)$. Hence the joint weight w of the pair $(\mathbf{u},\mathbf{v})$ is same as the weight of the polynomial $e(X,\eta)$. Since the BCH distance of $h$ is $2\tau+1$, if $w$ is at most $\tau$ then using Berlekamp algorithm (Theorem~\ref{thm:berlekamp-recast}) we can compute the polynomial $e$ and therefore the vectors $\mathbf{u}$ and $\mathbf{v}$. Applying $V^{\dagger}_{\mathbf{v}}U^{\dagger}_{\mathbf{u}}$ on $\ket{\psi}$ we recover the original codeword.
\end{IEEEproof}

\section{Density of numbers that divide $p^t +1$}

Let $f_p(x)$ denote the number of positive integers less than $x$ such
that $n - 1 = p^t \mod n$ for some $t$, i.e.
\begin{eqnarray*}
  f_p(x) & = & \# \{ n \leq x | \exists t \ n - 1 = p^t \mod n \}\\
         & = & \# \{ n \leq x | \exists t \ p^t  = -1 \mod n \}
\end{eqnarray*}

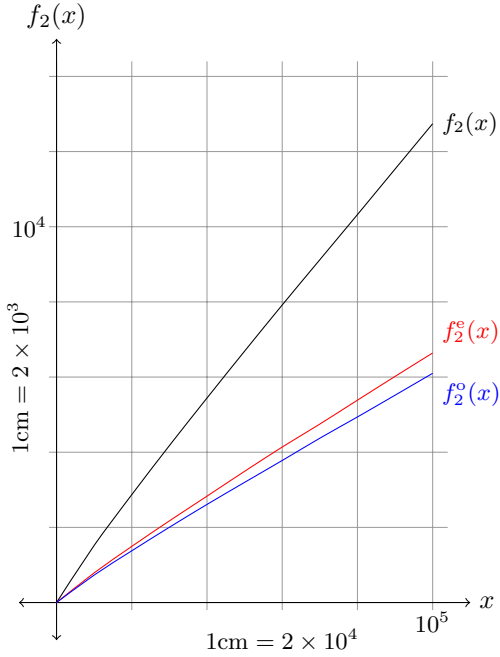
\begin{figure}
\begin{tikzpicture}
  \draw[very thin, color=black!40] (-0.1,-0.1) grid (5.2,7.2);
  \draw[<->] (-0.5,0) -- (5.5 ,0) node [right] {$x$};
  \draw[<->] (0,-0.5) -- (0,7.5) node [above] {$f_2(x)$};
  \draw [xscale=0.5,yscale=5] plot [smooth] file {density.total}
        node [right] {\small{$f_2(x)$}};
  \draw [color=red, xscale=0.5,yscale=5] plot [smooth] file {density.even}
        node [color=red,above right] {\small{$f_2^{\mathrm{e}}(x)$}};
  \draw [color=blue,xscale=0.5,yscale=5] plot [smooth] file {density.odd}
        node [color=blue,below right] {\small{$f_2^{\mathrm{o}}(x)$}};
  \path (5,0) node [below] {\small{$10^5$}};
  \path (0,5) node [left] {\small{$10^4$}};

  \node at (3,-0.5) {\small{$1\mathrm{cm} = 2 \times 10^4$}};
  \node at (-0.5,3) [rotate=90] {\small{$1\mathrm{cm} = 2 \times 10^3$}};
\end{tikzpicture}
\caption{Density of good lengths for characteristic 2}
\label{fig:plot-of-frobenius}
\end{figure}

For $p=2$ we have $f_2(100,000) = 12741$ of which $6641$ lengths
divide $2^t +1$ for an even exponent $t$ and the rest 66100 for an odd
exponent $t$.  Figure~\ref{fig:plot-of-frobenius} gives plot of
$f_2(x)$ for $x$ in the range $[0,10^5]$ and Table~\ref{tab:value_f_2}
gives some explicit values.

\begin{table}[tbh]
\begin{center}
\begin{tabular}{|l|l|l|l|}
\hline
$x$ & $f_2(x)$ & $f_2^{e}(x)$ & $f_2^{o}(x)$\\
\hline
10       & 2     & 1    & 1    \\
100      & 23    & 11   & 12   \\ 
1,000    & 189   & 101  & 88   \\
10,000   & 1521  & 790  & 731  \\
100,000  & 12741 & 6641 & 6100 \\
\hline
\end{tabular}
\end{center}
\caption{Values of $f_2(x)$}\label{tab:value_f_2}
\end{table}

We believe that the density is of $f_p(x)$ the form
$\frac{x}{\sqrt{\log{x}}}$, although the plots look linear. We show a
weaker lower bound

\begin{claim}
  For any prime $p$, $f_p(x) \geq c_p\frac{x}{\log{x}}$, where $c_p$
  is a constant that depends only on $p$.
\end{claim}
\begin{proof}
  Fix a characteristic $p$ of the base field. Let us estimate only the
  prime lengths $n$ that are good, i.e. primes $n$ such that $p^t = -1
  \mod n$. When $n$ is also a prime, if $p$ has an even order say
  $2\ell$ in the group $\mathbb{Z}/n\mathbb{Z}^*$ then $p^\ell = -1$
  and hence $n$ is good. In particular, if $p$ is a quadratic
  non-residue modulo $n$ then $n$ is good.  First consider the case
  when $p=2$. Using quadratic reciprocity, we have $2$ is a quadratic
  non-residue if and only if $n \equiv 3 \textrm { or } 5 \mod
  8$. Therefore, the density $f_2(X)$ is at least the density of
  primes in the arithmetic progression $3 \mod 8$ (or for that matter
  $5 \mod 8$). We can now use the density version of Dirichlet's
  theorem on prime numbers in AP.

  On the other hand if $p$ is odd then again using quadratic
  reciprocity we have

  \begin{equation}\label{eqn-reciprocity}
    \left(\frac{p}{n}\right) = (-1)^{\frac{n-1}{2}\frac{p-1}{2}}
    \left(\frac{n}{p}\right).
  \end{equation}

  Pick any quadratic non-residue $1 \leq \beta \leq p -1$ modulo
  $p$. Then, from Equation~\ref{eqn-reciprocity}, $n$ is good if it
  simultaneously satisfies the equations.

  \begin{equation}\label{eqn-prime-ap}
    n \equiv \left\{
      \begin{array}{l}
        \beta \mod p\\
        1 \mod 4
      \end{array}\right.
  \end{equation}

  Using Chinese reminder theorem, we can find an element $n_0$
  satisfying Equation~\ref{eqn-prime-ap} and coprime to $4p$ such that
  $1 \leq n_0 \leq 4p$. Therefore, all $n$ that is $n_0$ modulo $4p$
  are good. The result then follows by using the density version of
  Dirichlet's theorem.
\end{proof}

\section{Conclusion}
In this paper, we studied cyclic stabiliser codes of length dividing
$p^t +1$ over $\mathbb{F}_p$. It is natural to ask whether
the construction can be generalised for arbitrary code length. For 
higher degree extensions the gap between actual and BCH
distance could be significant. Therefore, it would be interesting to
find a better lower bound and in particular to know whether
Berlekamp like algorithms can be used to decode up to that
bound. Unlike previous definition of cyclicity, our definition
is applicable to non-stabiliser codes as well. An open problem is to
construct cyclic non-stabiliser codes.


\bibliography{bibdata} 
\bibliographystyle{abbrv}

\newpage
\appendix
\section*{Proof of Proposition~\ref{prop:stab-cyclic}} 

\section*{Proof of Lemma \ref{lem:cyclic-linear}}

\section*{Frobenius involution}
The Frobenius involution satisfies few properties
which are crucial in many of our proofs.

\begin{proposition}\label{prop:frobenius-involution}
  Let $\sigma$ denote the Frobenius involution on the ring of
  polynomials over the quadratic extension
  $\mathbb{F}_p(\eta)/\mathbb{F}_p$. Then
  \begin{enumerate}
  \item Let $\eta'$ be the conjugate $\eta^p$ of $\eta$ then for
    polynomials $a(X)$ and $b(X)$ over $\mathbb{F}_p$ we have
    $\sigma(a + \eta b) = a + \eta' b$.
  \item Any $a(X, \eta)$ in $\mathbb{F}_p(\eta)$ is a polynomial over
    $\mathbb{F}_p$ if and only if $\sigma(a) = a$.
  \item Let $a$, $b$ and $g$ be polynomials over $\mathbb{F}_p(\eta)$
    then $a = b \mod g$ if and only if $\sigma(a) = \sigma(b) \mod
    \sigma(g)$.
  \end{enumerate}
\end{proposition}







\section*{Explicit examples of linear Frobenius codes}
We now demonstrate our construction for the case when the
characteristic $p$ of the underlying finite field is 2. 
 Tables~\ref{tab:explicit-examples-2} and \ref{tab:explicit-examples-3} give some explicit codes
for all lengths $n$ less than 100 which is a factor of $2^{dm}+1$ for
some $m$ where $d$ is either $2$ or $3$. When $d=2$ the codes are all linear
and for $d=3$ they are non-linear. Recall that for a $t$-Frobenius code with
canonical factorisation $g \cdot h$, it is necessary for both $X+1$
and $X-1$ to divide $g$. However, since $1$ and $-1$ are the same in
$\mathbb{F}_2$, the polynomial $g$ needs to have only the factor
$X-1$. The notation used in the tables are the following: Let $\beta$
denote any fixed primitive $n$-th root of unity. Roots of any degree
$l$ irreducible factor of $X^n-1$ over $\mathbb{F}_q$ are exactly
$\beta^{kq^0}\ldots \beta^{kq^{l-1}}$ for some $k$, where $l$ is the
smallest positive integer such that $kq^l = k \mod n$. Call this
factor $f_{q,k}$. Let the polynomials $g_k$ and $h_k$ in the tables
denote $f_{2,k}$ and $f_{2^d,k}$ respectively. Notice that $g_0=X-1$ and
in case of $d=2$, for any $k \neq 0$, $g_k=h_k h_{2k}$ where $\sigma(h_k)=h_{2k}$.  Similarly in case of $d=3$ if degree of $g_k$ is divisible by $3$ then 
$g_k=h_k h_{2k} h_{3k}$. The
distance given in these tables is the BCH distance.  The actual distance
can be larger.

\begin{table*}[tbh]
  \begin{center}
  \begin{tabular}{|*{5}{>{\high $}l<{$}|}}
\hline
m & \multicolumn{2}{c|}{Canonical factors} & \high \text{Roots of } h & \text{Code}\\
\hline
1 & g_{0} & h_{2} & \beta^{2},\ldots,\beta^{3} & [[5,1,3]] \\ 
\hline
3 & g_{0} & h_{2} & \beta^{5},\ldots,\beta^{8} & [[13,1,5]] \\ 
\hline
\multirow{2}{*}{2}
  & g_{0} & h_{2}h_{6} & \beta^{6},\ldots,\beta^{11} & [[17,1,7]] \\ 
\cline{2-5}
  & g_{0}g_{1} & h_{6} & \beta^{6},\ldots,\beta^{7} & [[17,9,3]] \\ 
\hline
\multirow{2}{*}{5}
  & g_{0} & h_{1}h_{5} & \beta^{4},\ldots,\beta^{6} & [[25,1,4]] \\ 
\cline{2-5}
  & g_{0}g_{5} & h_{2} & \beta^{2},\ldots,\beta^{3} & [[25,5,3]] \\ 
\hline
7 & g_{0} & h_{1} & \beta^{4},\ldots,\beta^{7} & [[29,1,5]] \\ 
\hline
9 & g_{0} & h_{1} & \beta^{9},\ldots,\beta^{12} & [[37,1,5]] \\ 
\hline
\multirow{2}{*}{2}
  & g_{0} & h_{1}h_{6} & \beta^{14},\ldots,\beta^{19} & [[41,1,7]] \\ 
\cline{2-5}
  & g_{0}g_{1} & h_{3} & \beta^{11},\ldots,\beta^{13} & [[41,21,4]] \\ 
\hline
13 & g_{0} & h_{2} & \beta^{18},\ldots,\beta^{23} & [[53,1,7]] \\ 
\hline
15 & g_{0} & h_{2} & \beta^{28},\ldots,\beta^{33} & [[61,1,7]] \\ 
\hline
\multirow{4}{*}{3}
  & g_{0}g_{1} & h_{6}h_{7}h_{10}h_{22}h_{26} & \beta^{22},\ldots,\beta^{28} & [[65,13,8]] \\ 
\cline{2-5}
  & g_{0}g_{11}g_{13} & h_{2}h_{6}h_{9}h_{10} & \beta^{29},\ldots,\beta^{36} & [[65,17,9]] \\ 
\cline{2-5}
  & g_{0}g_{7}g_{11}g_{13} & h_{2}h_{6}h_{10} & \beta^{30},\ldots,\beta^{35} & [[65,29,7]] \\ 
\cline{2-5}
  & g_{0}g_{5}g_{7}g_{11}g_{13} & h_{2}h_{6} & \beta^{31},\ldots,\beta^{34} & [[65,41,5]] \\ 
\cline{2-5}
  & g_{0}g_{1}g_{3}g_{5}g_{7}g_{13} & h_{22} & \beta^{22},\ldots,\beta^{23} & [[65,53,3]] \\ 
\hline
\multirow{2}{*}{12}
  & g_{0} & h_{1}h_{7} & \beta^{33},\ldots,\beta^{40} & [[97,1,9]] \\ 
\cline{2-5}
  & g_{0}g_{1} & h_{7} & \beta^{37},\ldots,\beta^{40} & [[97,49,5]] \\ 
\hline
  \end{tabular}
  \end{center}
  \caption{Linear cyclic stabiliser codes of length dividing $4^m +1$ over $\mathbb{F}_2$}
  \label{tab:explicit-examples-2}
\end{table*}

\begin{table*}[tbh]
  \begin{center}
  \begin{tabular}{|*{5}{>{\high $}l<{$}|}}
\hline
m & \multicolumn{2}{c|}{Canonical factors} & \high \text{Roots of } h & \text{Code}\\
\hline
1 & g_{0}g_{3} & h_{4} & \beta^{4},\ldots,\beta^{5} & [[9,3,3]] \\
\hline
2 & g_{0} & h_{4} & \beta^{6},\ldots,\beta^{7} & [[13,1,3]] \\
\hline
3 & g_{0} & h_{4} & \beta^{9},\ldots,\beta^{10} & [[19,1,3]] \\
\hline
3 & g_{0}g_{9} & h_{4}h_{12} & \beta^{12},\ldots,\beta^{15} & [[27,3,5]] \\
\hline
3 & g_{0}g_{9}g_{1} & h_{12} & \beta^{15},\ldots,\beta^{15} & [[27,21,2]] \\
\hline
3 & g_{0}g_{9}g_{3} & h_{4} & \beta^{22},\ldots,\beta^{23} & [[27,9,3]] \\
\hline
6 & g_{0} & h_{4} & \beta^{32},\ldots,\beta^{34} & [[37,1,4]] \\
\hline
3 & g_{0}g_{19} & h_{4}h_{12}h_{20} & \beta^{25},\ldots,\beta^{32} & [[57,3,9]] \\
\hline
3 & g_{0}g_{19}g_{5} & h_{4}h_{12} & \beta^{27},\ldots,\beta^{30} & [[57,21,5]] \\
\hline
3 & g_{0}g_{19}g_{3}g_{5} & h_{4} & \beta^{28},\ldots,\beta^{29} & [[57,39,3]] \\
\hline
10 & g_{0} & h_{4} & \beta^{29},\ldots,\beta^{32} & [[61,1,5]] \\
\hline
2 & g_{0}g_{13} & h_{4}h_{12}h_{20}h_{28}h_{44} & \beta^{27},\ldots,\beta^{38} & [[65,5,13]] \\
\hline
2 & g_{0}g_{13}g_{11} & h_{4}h_{12}h_{20}h_{28} & \beta^{28},\ldots,\beta^{37} & [[65,17,11]] \\
\hline
2 & g_{0}g_{13}g_{1}g_{3}g_{5}g_{11} & h_{28} & \beta^{36},\ldots,\beta^{37} & [[65,53,3]] \\
\hline
2 & g_{0}g_{13}g_{7}g_{11} & h_{4}h_{12}h_{20} & \beta^{30},\ldots,\beta^{35} & [[65,29,7]] \\
\hline
2 & g_{0}g_{13}g_{5}g_{7}g_{11} & h_{4}h_{12} & \beta^{31},\ldots,\beta^{34} & [[65,41,5]] \\
\hline
2 & g_{0}g_{13}g_{3}g_{5}g_{7}g_{11} & h_{4} & \beta^{32},\ldots,\beta^{33} & [[65,53,3]] \\
\hline
11 & g_{0} & h_{4} & \beta^{31},\ldots,\beta^{36} & [[67,1,7]] \\
\hline
9 & g_{0}g_{27}g_{3} & h_{1}h_{36} & \beta^{44},\ldots,\beta^{46} & [[81,21,4]] \\
\hline
9 & g_{0}g_{27}g_{1}g_{3} & h_{36} & \beta^{45},\ldots,\beta^{45} & [[81,75,2]] \\
\hline
9 & g_{0}g_{27}g_{9} & h_{4}h_{12} & \beta^{66},\ldots,\beta^{69} & [[81,9,5]] \\
\hline
9 & g_{0}g_{27}g_{3}g_{9} & h_{4} & \beta^{76},\ldots,\beta^{77} & [[81,27,3]] \\
\hline
8 & g_{0} & h_{2}h_{20} & \beta^{51},\ldots,\beta^{55} & [[97,1,6]] \\
\hline
8 & g_{0}g_{1} & h_{20} & \beta^{77},\ldots,\beta^{78} & [[97,49,3]] \\
\hline
8 & g_{0}g_{5} & h_{4} & \beta^{48},\ldots,\beta^{49} & [[97,49,3]] \\
\hline
5 & g_{0}g_{3}g_{9}g_{15}g_{33}g_{1}g_{5} & h_{44} & \beta^{55},\ldots,\beta^{55} & [[99,93,2]] \\
\hline
5 & g_{0}g_{3}g_{9}g_{15}g_{33}g_{1}g_{11} & h_{5} & \beta^{85},\ldots,\beta^{86} & [[99,69,3]] \\
\hline
5 & g_{0}g_{3}g_{9}g_{15}g_{33}g_{5}g_{11} & h_{4} & \beta^{67},\ldots,\beta^{68} & [[99,69,3]] \\
\hline
\end{tabular}
\end{center}
  \caption{Non-linear cyclic stabiliser codes of length dividing $8^m +1$ over $\mathbb{F}_2$}
  \label{tab:explicit-examples-3}
\end{table*}

\end{document}